\documentclass{webofc}

\usepackage[varg]{txfonts}   
\usepackage{hyperref}
\usepackage{url}
\hypersetup{colorlinks=true,citecolor=blue,urlcolor=blue,linkcolor=blue}
\begin{document}
\title{A new effective theory for stochastic relativistic hydrodynamics}
%
%

\author{\firstname{Nicki} \lastname{Mullins}\inst{1}\fnsep\thanks{\email{nickim2@illinois.edu}} \and
        \firstname{Mauricio} \lastname{Hippert}\inst{2}\fnsep\thanks{\email{hippert@cbpf.br}} \and
        \firstname{Jorge} \lastname{Noronha}\inst{1}\fnsep\thanks{\email{jn0508@illinois.edu}}
}

\institute{Illinois Center for Advanced Studies of the Universe\\ Department of Physics, 
University of Illinois at Urbana-Champaign, Urbana, IL 61801, USA
\and
           Centro Brasileiro de Pesquisas Físicas, Rio de Janeiro, RJ, 22290-180, Brasil
          }

\abstract{Thermal fluctuations are a fundamental feature of dissipative systems that are essential for understanding physics near the expected critical point of QCD and in small systems. 
When such fluctuations are modeled naively in relativistic systems, strange features can appear such as negative self-correlation functions. 
We construct an effective theory for nonlinear stochastic relativistic hydrodynamics that ensure a well-posed mathematical formulation. 
Using Crooks fluctuation theorem, we derive a symmetry of the effective action that incorporates fluctuations through a suitable free energy functional. 
For divergence type theories, the action can then be fully specified using a single vector generating current. 
The equations of motion obtained using this procedure are guaranteed to be flux conservative and symmetric hyperbolic when the dynamics is causal. 
This ensures that these equations are well-posed (for suitable initial data) and are in a form that can easily be simulated, including with Metropolis techniques.
}

\maketitle

\section{Introduction}

The quark-gluon plasma (QGP) formed in heavy-ion collisions is well described as a viscous relativistic fluid. The modeling of such matter typically involves solve standard partial differential equations for the evolution of one-point functions of various fields that constitute the conserved energy-momentum tensor $T^{\mu\nu}$ and any conserved currents $J^{\mu}$. However, thermal systems also experience fluctuations that are encoded in higher-point functions for these dynamical variables. In typical many-body systems these fluctuations are small, however it is known that these fluctuations grow increasingly important near the critical point of a phase transition and in systems with few degrees of freedom, two scenarios relevant for the study of the QGP. These fluctuations can be modeled via stochastic relativistic hydrodynamics.

There has thus been significant recent interest in the development of mathematically consistent and practically solvable formulations of stochastic relativistic hydrodynamics \cite{Basar:2024srd}. In this work, we present a new effective theory for constructing such theories which has a number of desirable properties. In particular, the resulting hydrodynamic theories will always be of divergence type \cite{Geroch:1990bw}, which ensures that the non-fluctuating equations will be flux conservative. This structure allows for the simple imposition of relativistic causality conditions, a necessity for stable evolution around the equilibrium state in Lorentz invariant systems \cite{Gavassino:2021owo}. The second law of thermodynamics holds exactly, which ensures that there exists a sensible steady state and the probability distribution around such a state can be determined by the entropy and external work. 

\section{Stochastic relativistic hydrodynamics}

Hydrodynamics is an effective theory for the long-wavelength, late-time dynamics of conserved quantities. This is typically realized through a set of conservation equations for the one point correlation function of the energy momentum tensor, $\langle T^{\mu\nu} \rangle$ and any conserved currents $\langle J^{\mu} \rangle$. At this level, the dynamics is fully determined by a set of partial differential equations, $D_{AB} \phi^B = 0$, where $D_{AB}$ is some differential operator, $\phi^A$ is the list of dynamical variables, and capital letter indices run over the space of dynamical variables. 

However, in many situations we are interested in not only the evolution of one-point functions, but also in fluctuations of the conserved quantities. Such fluctuations can be introduced by including a stochastic source in the equations of motion, $D_{AB} \phi^B - \xi_A$, where $\xi_A$ is a random variable drawn from some probability distribution $P[\xi]$. When such a stochastic source is included, the solution to the equations of motion is no longer deterministic, instead there is an ensemble of solutions corresponding to different realizations of the noise $\xi_A$. The system can then be characterized by a partition function
\begin{equation}
    Z = \int D \phi D \xi \,\delta \left( D_{AB} \phi^B - \xi_A \right) P[\xi] . 
\end{equation}
In the case of linear systems, this can be written as a path integral over an effective action with doubled degrees of freedom \cite{Martin:1973zz}. In this work, we present an effective theory approach that allows us to construct such effective actions from the ground up for causal and stable hydrodynamic systems. 

\section{Effective theory from Crooks theorem}

The effective theory considered in this work \cite{Mullins:2025vqa} uses two principles: divergence type hydrodynamics \cite{Geroch:1990bw} and Crooks fluctuation theorem \cite{Crooks_1999}. 

\subsection{Divergence type hydrodynamics}

Consider an ideal fluid system described by $\alpha = \mu / T$ for some chemical potential $\mu$ and temperature $T$, and $\beta^{\mu} = u^{\mu} / T$ for fluid velocity $u^{\mu}$. Then, the first law of thermodynamics can be written in the form $dX^{\mu} = J^{\mu} d\alpha + T^{\mu\nu} d\beta_{\nu}$,
where $X^{\mu} = P \beta^{\mu}$. This implies that $X^{\mu}$ acts as a generating current such that taking a derivative with respect to a suitable dynamical variable recovers the conserved currents. 

Divergence type hydrodynamics postulates that dissipative models can be constructed along this same idea. The generating current $X^{\mu}$ is expanded to include non-equilibrium terms such that $J^{\mu} = \partial X^{\mu} / \partial \alpha$ and $T^{\mu\nu} = \partial X^{\mu} / \partial \beta_{\nu}$, 
including the proper dissipative corrections. The full equations of motion are taken to have the form 
\begin{equation}
    \nabla_{\mu} \frac{\partial X^{\mu}}{\partial \phi^A} = I_A(\phi) , 
\end{equation}
where $\phi^A = \{ \alpha, \beta_{\mu}, \phi^a \}$ for some set of dissipative variables $\phi^a$, and $I_A$ is some source term. These equations of motion are always flux-conservative and causality conditions are fully determined by the form of the generating current. The corresponding entropy current is then given by 
\begin{equation}
    s^{\mu} = X^{\mu} - \phi^A \frac{\partial X^{\mu}}{\partial \phi^A} \quad \rightarrow \quad \sigma \equiv \nabla_{\mu} s^{\mu} = \phi^A I_A .
\end{equation}
So, the second law of thermodynamics can be imposed by choosing a sensible source term $I_A$. The leading order choice is $I_A = \Xi_{AB} \phi^B$, where $\Xi_{AB}$ is positive semi-definite with support only in the dissipative subspace. 

\subsection{Crooks fluctuation theorem}

So far we have not included any information about thermal fluctuations. This can be done by demanding that the stochastic path integral obeys Crooks fluctuation theorem \cite{Crooks_1999}. This can be expressed schematically as $P[A \rightarrow B] = P_{\Theta}[B \rightarrow A] e^{\omega}$,
where $P[A \rightarrow B]$ denotes the transition probability from state $A$ to state $B$, $\Theta$ denotes the action of the relevant discrete symmetries, and $\omega = \int d^4x \sqrt{-g} \sigma$ is the total entropy produced during the transition from states $A$ to $B$. The transition probability can be related to the path integral by 
\begin{equation}
    P[A \rightarrow B] \sim \int_A^B \mathcal{D} \phi \mathcal{D} \Bar{\phi} e^{i \int d^4x \sqrt{-g} \mathcal{L}[\phi,\bar{\phi}]} , 
\end{equation}
for some effective Lagrangian $\mathcal{L}[\phi,\bar{\phi}]$. Then, we find that consistency with Crooks fluctuation theorem requires this Lagrangian to obey
\begin{equation} \label{Eq:symmetry}
    \mathcal{L}[\Theta \phi, \Theta \bar{\phi} - i \Theta \phi] = \mathcal{L}[\phi,\bar{\phi}] - i \sigma .
\end{equation}
Imposing this symmetry on the effective action will then enforce non-linear fluctuation-dissipation relations \cite{Mullins:2025vqa}. 

\subsection{Hydrodynamic effective theory}

We now want an effective action that corresponds to a divergence type theory when fluctuations are neglected, but also incorporates nonlinear fluctuation-dissipation relations through the symmetry of Eq.~\eqref{Eq:symmetry}. Up to leading order, this results in the Lagrangian
\begin{equation}
    \mathcal{L} = -\bar{\phi}^A \nabla_{\mu} \frac{\partial X^{\mu}}{\partial \phi^A} + i \zeta^{ab}_0(\phi) \bar{\phi}_a \left( \bar{\phi}_b + i \phi_b \right) .
\end{equation}
Here, $\zeta^{ab}_0(\phi)$ is an even function of the fields and has support only on the dissipative subspace of dynamical variables so that no entropy can be produced in equilibrium. This action can be fully specified using a generating current $X^{\mu}$, which is determined by the form of the conserved currents, and the ``dissipative potential" $\zeta^{ab}_0(\phi)$, which is determined by the entropy production. The equations of motion always have a Lyapunov functional which ensures stability against thermal fluctuations \cite{Gavassino:2021kjm}, and the distribution of the resulting thermal noise does not depend on the choice of spacetime foliation \cite{Mullins:2023tjg, Mullins:2023ott}.

As an example, we consider the case of relativistic diffusion. This corresponds to the dynamics of a single conserved current, so the equilibrium degree of freedom is $\alpha = \mu / T$. We also consider a dissipative transverse vector $j^{\mu}$, so that the conserved current is $J^{\mu} = n u^{\mu} + j^{\mu}$ with constant background velocity $u^{\mu}$. The generating current is then $X^{\mu} = P(\alpha, j^2) \beta^{\mu} + \alpha j^{\mu}$.
This obeys $\partial X^{\mu} / \partial \alpha = n u^{\mu} + j^{\mu}$ as expected. The corresponding Lagrangian at leading order is given by 
\begin{equation}
    \mathcal{L} = - \bar{\alpha} \nabla_{\mu} J^{\mu} - \bar{j}_{\nu} \nabla_{\mu} \left( 2 \frac{\partial P}{\partial j^2} \beta^{\mu} j^{\nu} + \alpha \Delta^{\mu\nu} \right) + \frac{i}{\kappa T} \bar{j}_{\mu} \left( \bar{j}^{\mu} + i j^{\mu} \right) .
\end{equation}
This describes both the conservation law and a fluctuating relaxation equation for the dissipative vector $j^{\mu}$. 

\section{Conclusions}

The formalism discussed in this work provides a systematic way to construct effective actions for stochastic relativistic hydrodynamics for which it is easy to impose causality constraints. The existence of a covariant Lyapunov functional ensures stability of the steady state against thermal fluctuations and independence on the choice of spacetime foliation \cite{Mullins:2023ott, Mullins:2023tjg}. It would be interesting to see how this approach can be used to study hydrodynamic theories constructed using the gradient expansion as there are challenges associated with incorporating fluctuations in causal versions of such theories \cite{Mullins:2023ott, Gavassino:2024vyu}.

\section*{Acknowledgements}
JN and NM are partly supported by the U.S. Department of Energy, Office of Science, Office for Nuclear Physics
under Award No. DE-SC0023861. 
M.H. was supported in part by the Universidade Estadual do Rio de Janeiro through the Programa de Apoio à Docência (PAPD), and by the Brazilian National Council for Scientific and Technological Development (CNPq) under process No. 313638/2025-0.

\bibliography{references}

\end{document}